\documentclass[sigconf]{acmart}

\usepackage{graphicx}
\usepackage{bm}
\usepackage{setspace}
\usepackage{amsmath,amsfonts,amsthm}
\usepackage{booktabs}

\usepackage{textcomp}
\usepackage{xcolor}
\usepackage[font=footnotesize]{subfig}
\usepackage{multirow}
\usepackage{float}

\setcopyright{acmcopyright}

\copyrightyear{2020}
\acmYear{2020}
\setcopyright{acmcopyright}\acmConference[SIGIR '20]{Proceedings of the 43rd International ACM SIGIR Conference on Research and Development in Information Retrieval}{July 25--30, 2020}{Virtual Event, China}
\acmPrice{15.00}
\acmDOI{10.1145/3397271.3401304}
\acmISBN{978-1-4503-8016-4/20/07}

\AtBeginDocument{%
  \providecommand\BibTeX{{%
    \normalfont B\kern-0.5em{\scshape i\kern-0.25em b}\kern-0.8em\TeX}}}

\begin{document}
\fancyhead{}
\title{TFNet: Multi-Semantic Feature Interaction for CTR Prediction}

\author[Shu Wu, Feng Yu, Xueli Yu, Qiang Liu, Liang Wang, Tieniu Tan, Jie Shao, Fan Huang]{Shu Wu$^{1,2,*}$, Feng Yu$^{1,2,*}$, Xueli Yu$^1$, Qiang Liu$^{3,4}$, Liang Wang$^{1,2}$, Tieniu Tan$^{1,2}$, Jie Shao$^5$ and Fan Huang$^5$}

\makeatletter
\def\authornotetext#1{
	\g@addto@macro\@authornotes{%
	\stepcounter{footnote}\footnotetext{#1}}%
}
\makeatother

\authornotetext{The first two authors contributed equally to this work.}
\authornotetext{This work is supported by National Key Research and Development Program (2018YFB1402600).}

\affiliation{%
	\institution{$^1$Center for Research on Intelligent Perception and Computing, Institute of Automation, Chinese Academy of Sciences}
	\institution{$^2$University of Chinese Academy of Sciences \qquad 
	$^3$RealAI \qquad $^4$Tsinghua University \qquad $^5$Tencent }
}

\email{{feng.yu,shu.wu,wangliang,tnt}@nlpr.ia.ac.cn, xueli.yu@cripac.ia.ac.cn, qiang.liu@realai.ai,{jshao,sinohuang}@tencent.com}

\def\authors{Shu Wu, Feng Yu, Xueli Yu, Qiang Liu, Liang Wang, Tieniu Tan, Jie Shao, Fan Huang}

\begin{abstract}
  The CTR (Click-Through Rate) prediction plays a central role in the domain of computational advertising and recommender systems. There exists several kinds of methods proposed in this field, such as Logistic Regression (LR), Factorization Machines (FM) and deep learning based methods like Wide\&Deep, Neural Factorization Machines (NFM) and DeepFM. However, such approaches generally use the vector-product of each pair of features, which have ignored the different semantic spaces of the feature interactions. In this paper, we propose a novel Tensor-based Feature interaction Network (TFNet) model, which introduces an operating tensor to elaborate feature interactions via multi-slice matrices in multiple semantic spaces. Extensive offline and online experiments show that TFNet: 1) outperforms the competitive compared methods on the typical Criteo and Avazu datasets; 2) achieves large improvement of revenue and click rate in online A/B tests in the largest Chinese App recommender system, Tencent MyApp. 
\end{abstract}

\maketitle

\section{Introduction}
The CTR prediction plays a central role in the domain of computational advertising and recommender systems, where an item could whether be recommended is decided by the probability of whether the user would click on it. Traditional methods to predict the probability of CTR are LR and FM \cite{rendle2012factorization}. However, these kinds of methods can not obtain the higher-order interaction of different features. 
Several deep learning methods have been proposed to this field in recent years, such as Wide\&Deep \cite{cheng2016wide}, NFM \cite{he2017neural} and DeepFM \cite{guo2017deepfm}. The general architecture of these methods is simply concatenating the first-order features and interactive second-order features, inputting them into the Multilayer Perceptron (MLP) to learn higher-order feature interactions and finally predicting the click-through rate. For instance, the Wide\&Deep \cite{cheng2016wide} model jointly trains wide linear models and deep neural networks. 

However, there exists a potential limitation in such methods, they all have ignored the different semantic spaces among the feature interactions. In other words, the previous works generally use the vector-product of each pair of features, considering them all in one semantic space. Because of the features' semantic diversities, different interactive features may be in different semantic spaces. For instance, in ads recommender systems, it is reasonable that the interactions of feature pair (user, ad) and (banner-position, ad) are in the different semantic spaces, where the former learns the effect of the user's preference on the ad while the latter represents the effect of cost paid by the advertiser on this ad. Therefore, learning such feature interactions via simple vector-product in just one semantic space is obviously insufficient. 

There are several fields of works which have utilized the semantic interactions, such as Natural Language Processing (NLP) \cite{socher2013recursive} and recommender systems \cite{liu2015cot,Shu2016Contextual,liu2015convolutional}. For example, the work in NLP \cite{socher2013recursive} introduces a tensor-based composition function to learn powerful meaning of pairs of words vectors so as to realize better interactions between each pair of words. Besides, in recommender systems field, an operating tensor is used to explore the semantic effects on the recommendation results \cite{liu2015cot}. 

Inspired by the works above, we propose a method named TFNet. We introduce an operating tensor to elaborate feature interactions via multi-slice matrices, by which we can acquire the difference in multiple semantic spaces. The whole procedures of our model are listed following: we firstly input the original features and embed them into dense embedding vectors, then introduce the operating tensor to carry out the tensor-based feature interactions between each pair of field vectors. Further, we input the interactive features and the embedding vectors into DNN separately to model higher-order feature interactions  and finally combine the original raw features to make the ultimate prediction. We also demonstrate that TFNet is a more general form of recent models in following sections. 

In summary, our main contributions are as follows: 1) We propose a novel TFNet model, introducing a tensor-based approach to learn feature interactions in multiple semantic spaces, which can capture the interactive mechanism of different features more sufficiently than the current deep learning based models. 2) Extensive offline and online experiments show that TFNet not only outperforms competitive compared methods on typical Criteo and Avazu datasets, but also achieves large improvement of revenue and click rate in online A/B tests in the largest Chinese App recommender systems.

\begin{figure*}[t]
\setlength{\abovecaptionskip}{0pt}  
\setlength{\belowcaptionskip}{0pt}  
   \begin{center}
   \vspace{-4mm}
   \includegraphics[width=0.7\textwidth]{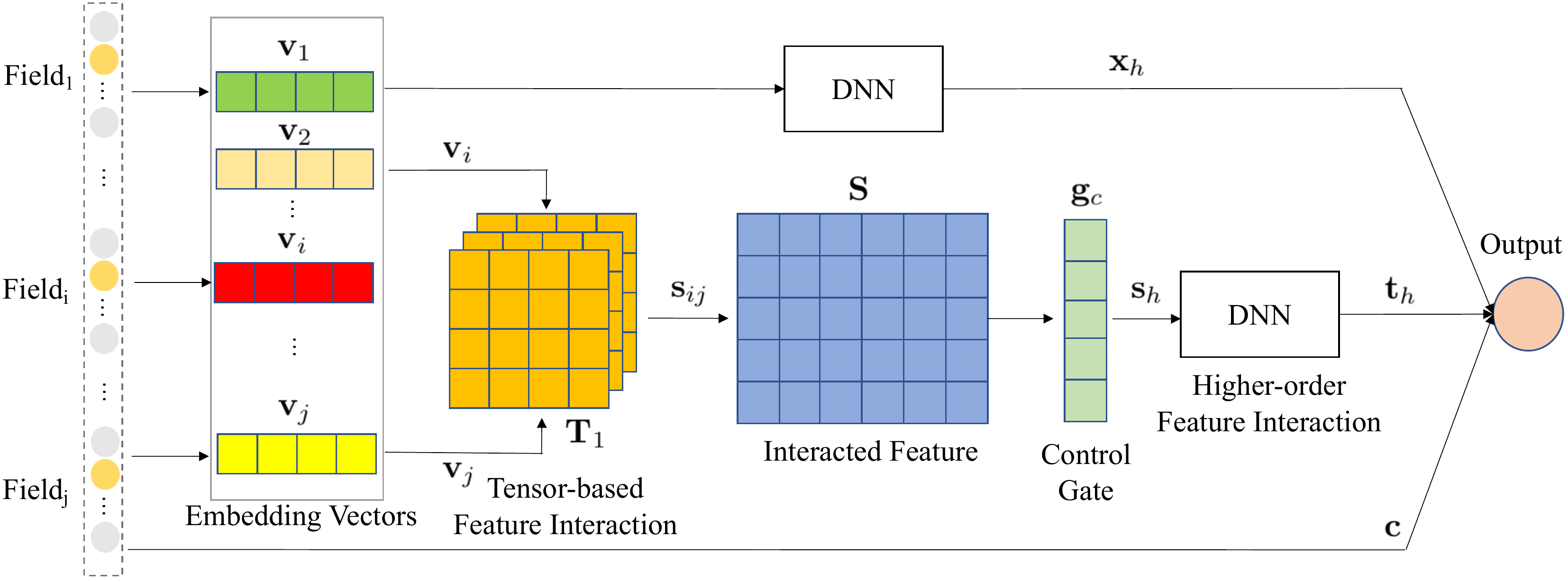}
   \end{center}
   \caption{The architecture of our proposed TFNet model, which consists of the following parts: sparse input layer, embedding layer, tensor-based feature interaction layer, higher-order feature interaction layer and the final output.}
   \label{TFnet1}
\end{figure*}

\section{Method}
The whole structure of TFNet model is shown in Figure \ref{TFnet1}. In this section, we detail the implement and how they work.

\subsection{Sparse Input and Embedding Layer}
The sparse input layer imports sparse original features, and low dimensional dense vectors of them are learned via the embedding layer. Supposing there are $n$ feature fields $Field_1,...,Field_n$ in an ad impression scenario, following previous works \cite{li2019fignn}, we input them into the sparse input and embedding layer, embedding each feature $Field_i$ as $\mathbf{v}_{i}\in \mathbb{R} ^{d}$, where ${d}$ is the dimension of $Field_i$'s embedding vectors. Therefore, we can obtain embedding vectors of all the fields as $\left\{ \mathbf{v}_1, \mathbf{v}_2, ..., \mathbf{v}_n\right\}$. The concatenation of them is defined as $\mathbf{x}_v$.

\subsection{Tensor-based Feature Interaction Layer} In this section, we aim to explore tensor-based feature interactions from the embedding of input features, capturing the feature interactions in different semantic spaces. 

\textbf{Tensor-based Semantic Interaction.} We bring in the operating tensor $\mathbf{T}_1$ to learn how to model the interaction between each pair of features $\mathbf{v}_i$ and $\mathbf{v}_j$, where $\mathbf{T}_1\in \mathbb{R} ^{d\times m\times d}$ is a third-order operating tensor and $m$ is the number of slices. Each slice $\mathbf{T}_1^{[i]}\in \mathbb{R} ^{d\times d}$ can represent an operation over paired-features in a semantic space. That is to say, the $\mathbf{T}_1$ stores several kinds of semantic operations spaces, for example, the user preference space and the advertiser preference space in the ads impression scenario. Based on the operating tensor $\mathbf{T}_1$ and the embeddings of input instances, we can generate an interactive feature $\mathbf{s}_{ij}\in \mathbb{R} ^{m}$ of $\mathbf{v}_i$ and $\mathbf{v}_j$ as follows,
\begin{equation}
\mathbf{s}_{ij} = \mathbf{v}_i^T{\mathbf{T}_1}\mathbf{v}_j~.
\label{equ:combination}
\end{equation}
Then we can concatenate all the interactive features to construct matrix $\mathbf{S}\in \mathbb{R} ^{q\times m}$, where $q=n*(n-1)/2~$. 

\begin{figure}
\setlength{\abovecaptionskip}{0pt}  
\setlength{\belowcaptionskip}{0pt}  
\centering
\includegraphics[width=0.35\textwidth]{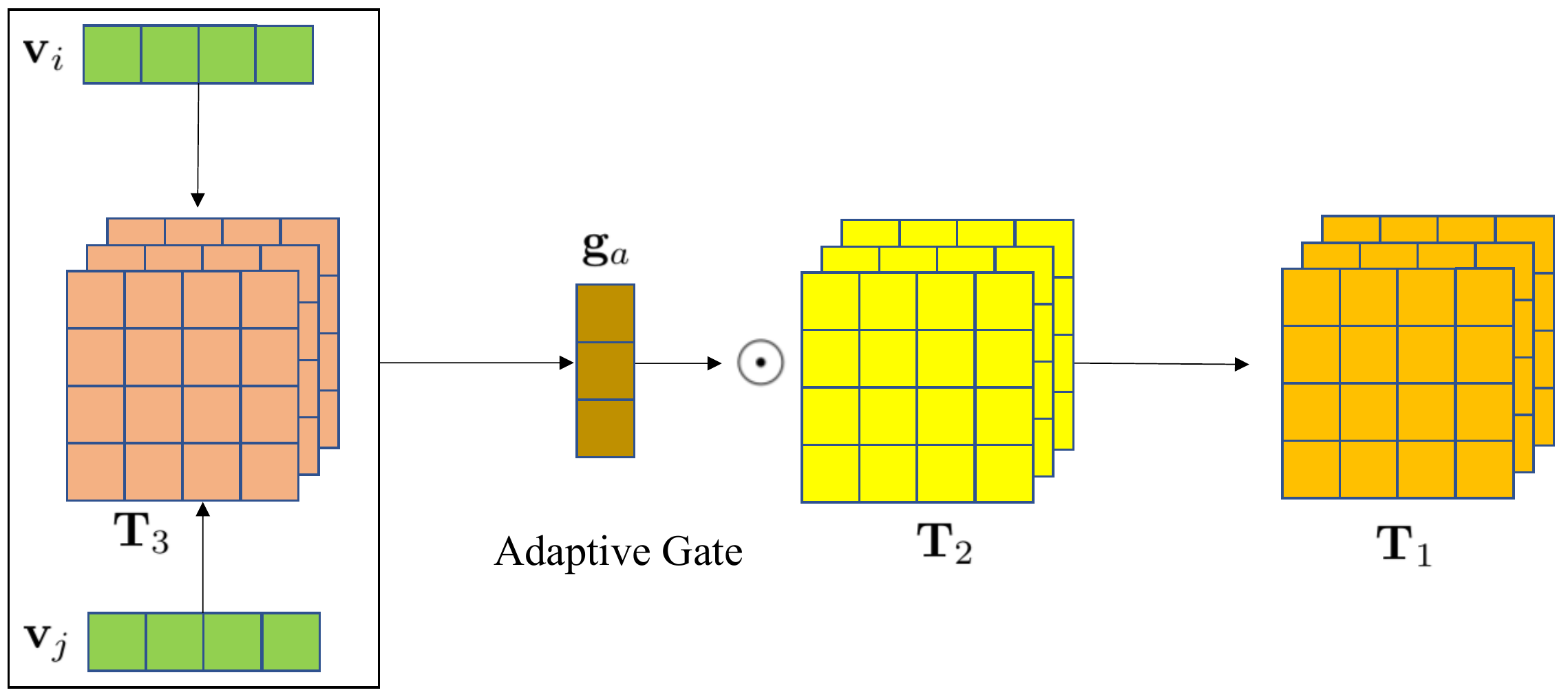}
\caption{The architecture of how to learn the weighted operating tensor $\mathbf{T}_1$.}
\label{TFNet2}
\end{figure}

\textbf{Adaptive Gate.} Because of the diversity of semantic spaces, modeling all feature interactions in different semantic spaces with the same weight  may be not sufficient. For example, when learning the interactions of feature pair (user, ad), the semantic space of user preference tends to be more important than the space of advertiser preference. Therefore, it is necessary to introduce an importance weight to the operating tensor, by which we can discriminate the importance of different semantic spaces during learning the feature interactions. In order to realize it, we introduce an adaptive gate with different importance weights to learn the operating tensor $\mathbf{T}_1$ . The architecture of how to learn the weighted $\mathbf{T}_1$ is shown in Figure \ref{TFNet2}. As is illustrated, $\mathbf{T}_1$ can be obtained via attention mechanism on meta-semantic operation tensor $\mathbf{T}_2\in \mathbb{R} ^{d\times m\times d}$ as 
\begin{equation}
\mathbf{T}_1=\mathbf{g}_a\odot \mathbf{T}_2~, 
\end{equation}
where $\odot$ means element-wise multiplication of a vector and a tensor, and the adaptive gate $\mathbf{g}_a\in \mathbb{R} ^{m}$ is the importance weight of meta-semantic operations for a specific pair of features interaction ($\mathbf{v}_i, \mathbf{v}_j$). Thus $\mathbf{T}_1^{[i]}$ of tensor $\mathbf{T}_1$ can be further interpreted as $\mathbf{T}_1^{[i]} = \mathbf{g}_a^{[i]}\mathbf{T}_2^{[i]}$. Moreover, to learn the attention score of $\mathbf{g}_a$, another operating tensor in the adaptive gate is introduced, which is the $\mathbf{T}_3\in \mathbb{R} ^{d\times m\times d}$, and the $\mathbf{g}_a$ is computed via $\mathbf{g}_a=softmax(\mathbf{v}_i^T{\mathbf{T}_3}\mathbf{v}_j)$. 

\textbf{Control Gate.}  Among all the generated interactive features above, it is noted that not all of them are useful for the target prediction, therefore, we need to select important features from all generated interactive features. Similar to \cite{wang2018interpret}, each new feature is associated with a control gate $\mathbf{g}_c\in \mathbb{R} ^{q}$ to decide whether the feature is useful for prediction. $\mathbf{g}_c$ should be non-negative and sparse, which is regularized by L1-Norm. Finally, the critical tensor-based interactive features can be obtained, defined as $\mathbf{s}_h = \mathbf{S}^T \mathbf{g}_c$.  

\subsection{Higher-order Feature Interaction Layer}
To model higher-order feature interactions, we further introduce a stack of fully connected layers as hidden layers $H_l$. Formally, the definition of fully connected layers is as follow:
\begin{align}
&\mathbf{z}_{i} = \sigma _i(\mathbf{W}_{i}\mathbf{z}_{i-1}+\mathbf{b}_{i})~,~i=1,2,...,l \\ 
&H_{l}(x) = \mathbf{z}_{l}(\mathbf{z}_{l-1}(...\mathbf{z}_{1}(x)))~,
\end{align}

\noindent where $l$ denotes the number of hidden layers, and the $\mathbf{W}_{i}$, $\mathbf{b}_{i}$ and $\mathbf{\sigma}_{i}$ is the weight matrix, bias vector and activation function of the $i$th layer. 
As is illustrated in Figure \ref{TFnet1}, there are two parts of higher-order feature interactions in the TFNet, including the tensor-based interactive features and the embedding vectors. 
For the tensor-based interactive features, we input the critical features $\mathbf{s}_h$ into the $l_1$-layer fully connected layers, obtaining the higher-order interactive features $\mathbf{t}_h = H_{l_1}(\mathbf{s}_{h})$. Besides, for the higher-order feature interactions of embedding vectors, the $l_2$-layer fully connected layers are constructed and the formulation of it is $\mathbf{x}_h = H_{l_2}(\mathbf{x}_{v})$.

\subsection{The Output Layer}
Finally, in the output layer, we concatenate three parts of features to make the ultimate prediction, including the original raw input features and the higher-order interactive features of embedding vectors and tensor-based critical features. The formulation of the final prediction score is as follow,
\begin{equation}
p = \sigma_s (\mathbf{w}^T\text{concat}(\mathbf{c}, \mathbf{x}_h, \mathbf{t}_h)+\mathbf{b}),
\end{equation}

\noindent where $p$ is the final prediction score of the TFNet model, $\mathbf{c}$ is the concatenation of original raw features. And the $\text{concat}$ function denotes the concatenation of the three parts. $\mathbf{w}$ and $\mathbf{b}$ are the model weight and the bias vector respectively. Furthermore, we use a non-linear activation function $\sigma _s$ to output the probabilities, which can be modified according to different tasks, for example, we use sigmoid for a classification task here. Given the label $y\in\{0,1\}$, the loss function is defined as the cross-entropy of prediction over the labels, formulated as 
\begin{equation}
\mathcal{L} = -\frac{1}{N} \sum_{i=1}^N y_{i}\log p_{i} + (1 - y_{i})\log (1 - p_{i}),
\end{equation}
where $N$ is the total number of training samples and $i$ indexes the training samples. Finally, the model can be trained in an end-to-end way efficiently using stochastic gradient descent algorithm.



\section{Experiments}
In this section, we evaluate the performance of the proposed TFNet model in both offline and online environments. We first describe the datasets and settings of the experiments, then report and analyze the experimental results. 

\subsection{Experiment Settings}
\textbf{Datasets.} A common application scenario of user click prediction is the task of click prediction of ad impression. The offline experiments are conducted on two typical and large real-world datasets.

$\mathbf{Criteo}\footnote{https://s3-eu-west-1.amazonaws.com/criteo-labs/dac.tar.gz}$ contains 45 million click records of ad impression. There are 13 continuous features and 26 categorical ones. Considering the higher volume and higher unbalance of data, we do negative down-sampling on the Criteo dataset as \cite{qu2018product}. $\mathbf{Avazu}\footnote{https://www.kaggle.com/c/avazu-ctr-prediction/data}$ contains 40 million click records of ad impression with 22 categorical features, such as site category, time, device type, banner position, etc. 


\textbf{Compared Methods.} Several representative methods are used for empirical comparison: (\romannumeral1) {FM} \cite{rendle2012factorization} , (\romannumeral2) {Wide\&Deep} \cite{cheng2016wide} , (\romannumeral3) {DeepFM} \cite{guo2017deepfm} , (\romannumeral4) {NFM} \cite{he2017neural} and (\romannumeral5) {AFM} \cite{xiao2017attentional}.

To make a fair comparison, the number of parameters of the proposed TFNet model is set to be approximately equal to that of most compared models. For the proposed TFNet model, the network structure of two hidden layers $H_{l_1}$,$H_{l_2}$ are 512-512, $d=45$, $m=4,6$ for Criteo dataset and Avazu dataset respectively. 

\textbf{Evaluation Metrics.} 
To evaluate the performance, we adopt \textbf{AUC} (Area Under ROC), and to estimate the Relative Improvement (RI) of online performance based on offline performance, \textbf{RI-AUC} is proposed to make good comparison between the proposed model and compared models \cite{chan2018convolutional}. 
\begin{equation}
\textit{RI}\text{-}\textit{AUC} = \dfrac {\textit{AUC}(model)-0.5}{\textit{AUC}(base)-0.5} -1~.
\end{equation}

For an online recommender business, there are two key evaluation metrics, \textbf{ARPU} (Average Revenue Per User) and \textbf{CTR}, which are defined as $\textit{ARPU} = R_t/N_u$, $\textit{CTR} = N_s/N_i$. ARPU is calculated in a standard time period, such as a day or a month. $R_t$ is the total revenue generated by all users during a time period. $N_u$ is the total number of users during that time period. In a recommender system, an impression means a view of a user to a recommended item. $N_i$ is the total number of impressions. $N_s$ is the number of impressions that are successfully recommended, which means users either click advertisements or download applications. Similar to \cite{qu2018product}, we can define relative improvement of ARPU and CTR, i.e., \textbf{RI-ARPU} and \textbf{RI-CTR} as follows ($X$ can be either ARPU or CTR),
\begin{equation}
\textit{RI}\text{-}\textit{X} = \dfrac {\textit{X}(model)-\textit{X}(base)}{\textit{X}(base)} *100\%~.
\end{equation}

\subsection{Offline Evaluations}
We conduct offline experiments to obtain thorough comparisons between our model and state-of-the-art models on two typical and large real-world datasets.\\

\noindent{\textbf{Results and Analysis}}

\noindent We first make ablation analysis of the higher-order feature interaction part of TFNet, which consists of the higher-order interactions of the embedding vectors and tensor-based interactive features. The analysis verifies that whether the higher-order part can be a complement of tensor-based ones in TFNet. Comparing TFNet-- and TFNet in the bottom part of Table \ref{compare}, higher-order interactions obtain extra 0.4\% relative AUC improvement, which demonstrates the mutual complementation of tensor-based and higher-order feature interaction. 
Table \ref{compare} also illustrates experimental results of other compared models on both Criteo and Avazu datasets. As is shown, the proposed TFNet model can gain prominent relative AUC improvement of around $2\%$ against compared models on both datasets. This verifies the effectiveness of the proposed method.

Compared with the baselines, the TFNet utilizes tensor-based interaction method to capture interactive features in different semantic spaces. As is shown in Table \ref{compare}, even without the higher-order feature interactions part of the TFNet model, the TFNet-- also outperforms the above models in both datasets, which verifies the effectiveness of the tensor-based approach.

\begin{table}[t]
\setlength{\abovecaptionskip}{0pt}  
\setlength{\belowcaptionskip}{0pt}  
\small
\centering
  \caption{Offline experimental results of compared methods on the Criteo and Avazu datasets. RI-AUC is the relative AUC improvement of the proposed TFNet model against all other models. {TFNet}-- means TFNet without higher-order feature interactions.}
\begin{tabular}{|c|c|c|c|c|}
\hline
\multirow{2}{*}{\textbf{Algorithm}} & \multicolumn{2}{c|}{\textbf{Avazu}} & \multicolumn{2}{c|}{\textbf{Criteo}} \\ \cline{2-5} 
                           & \textbf{AUC}        & \textbf{RI\text{-}AUC}   & \textbf{AUC}        & \textbf{RI\text{-}AUC}    \\ \hline
\textbf{FM}                & 77.67\%             & 3.22\%        & 78.93\%             & 3.39\%         \\ \hline
\textbf{AFM}               & 77.96\%             & 2.15\%        & 79.00\%             & 3.14\%         \\ \hline
\textbf{DeepFM}            & 77.99\%             & 2.04\%        & 79.34\%             & 1.94\%         \\ \hline
\textbf{NFM}               & 78.00\%             & 2.00\%        & 79.24\%             & 2.29\%         \\ \hline

\textbf{Wide\&Deep}        & 78.05\%             & 1.82\%        & 79.37\%             & 1.84\%         \\ \hline
\textbf{TFNet}--               & {78.43\%}    & {0.46\%}     & {79.79\%}    & {0.40\%}      \\ \hline
\textbf{TFNet (Ours)}               & \textbf{78.56\%}    & {--}     & \textbf{79.91\%}    & {--}      \\ \hline
\end{tabular}
\label{compare}
\end{table}

\noindent{\textbf{Hyper-Parameter Study}}

\noindent\textbf{Number of Slice $m$.}
As can be depicted in Table \ref{hyper}, when $m$ gradually increases, the performances of the proposed TFNet model on both datasets reach the peak and then decrease. It is worth mentioned that on the Criteo dataset, the TFNet achieves best AUC when $m=4$, while on the Avazu dataset the $m=6$. It may be likely that there are more feature fields in the Avazu dataset, which accordingly needs more slices of semantic spaces.

\textbf{Dimension $d$ of Embeddings.}  Table \ref{hyper} illustrates the AUC values of the proposed TFNet model on both datasets with different dimensions $d$ of latent embedding vectors. On both the Criteo and Avazu datasets, the TFNet model achieves the best AUC when dimension $d=45$. 
As the dimension $d$ continues to grow, the performance decreases due to overfitting. 

\begin{table}[]
\small
  \centering
  \caption{The impact of $d$ (dimension of embeddings) and $m$ (the number of slice of operating tensor) of the TFNet model on the Criteo and Avazu datasets.}
\begin{tabular}{|c|c|c|cccc}
\cline{1-3} \cline{5-7}
\textit{m} & Criteo  & Avazu   & \multicolumn{1}{c|}{} & \multicolumn{1}{c|}{\textit{d}} & \multicolumn{1}{c|}{Criteo}  & \multicolumn{1}{c|}{Avazu}   \\ \cline{1-3} \cline{5-7} 
1          & 0.79796 & 0.78461 & \multicolumn{1}{c|}{} & \multicolumn{1}{c|}{20}         & \multicolumn{1}{c|}{0.79840} & \multicolumn{1}{c|}{0.78382} \\ \cline{1-3} \cline{5-7} 
2          & 0.79835 & 0.78522 & \multicolumn{1}{c|}{} & \multicolumn{1}{c|}{25}         & \multicolumn{1}{c|}{0.79865} & \multicolumn{1}{c|}{0.78444} \\ \cline{1-3} \cline{5-7} 
3          & 0.79866 & 0.78529 & \multicolumn{1}{c|}{} & \multicolumn{1}{c|}{30}         & \multicolumn{1}{c|}{0.79866} & \multicolumn{1}{c|}{0.78499} \\ \cline{1-3} \cline{5-7} 
4          & \textbf{0.79910} & 0.78533 & \multicolumn{1}{c|}{} & \multicolumn{1}{c|}{35}         & \multicolumn{1}{c|}{0.79873} & \multicolumn{1}{c|}{0.78505} \\ \cline{1-3} \cline{5-7} 
5          & 0.79840 & 0.78537 & \multicolumn{1}{c|}{} & \multicolumn{1}{c|}{40}         & \multicolumn{1}{c|}{0.79897} & \multicolumn{1}{c|}{0.78536} \\ \cline{1-3} \cline{5-7} 
6          & 0.79844 & \textbf{0.78560} & \multicolumn{1}{c|}{} & \multicolumn{1}{c|}{45}         & \multicolumn{1}{c|}{\textbf{0.79910}} & \multicolumn{1}{c|}{\textbf{0.78561}} \\ \cline{1-3} \cline{5-7} 
7          & 0.79859 & 0.78544 & \multicolumn{1}{c|}{} & \multicolumn{1}{c|}{50}         & \multicolumn{1}{c|}{0.79877} & \multicolumn{1}{c|}{0.78545} \\ \cline{1-3} \cline{5-7} 
8          & 0.79886 & 0.78521 &                       &                                 &                              &                              \\ \cline{1-3}
\end{tabular}
\label{hyper}
\end{table}

\subsection{Online Evaluations}
We perform online A/B tests in Tencent MyApp\footnote{http://sj.qq.com/myapp/}, the largest Chinese App recommender system. The online baseline model is the Wide\&Deep method \cite{cheng2016wide}. The TFNet model is initialized by the latest 7-day CTR log data. During the following process of online evaluation, these two models are updated every two hours by using the same datasets: the training set includes CTR logs collected from the last hour to last 25 hours, and the validation set includes logs in the last hour.

\begin{figure}[t]
\setlength{\abovecaptionskip}{0pt}  
\setlength{\belowcaptionskip}{0pt}  
\centering
\small
\includegraphics[scale=0.4]{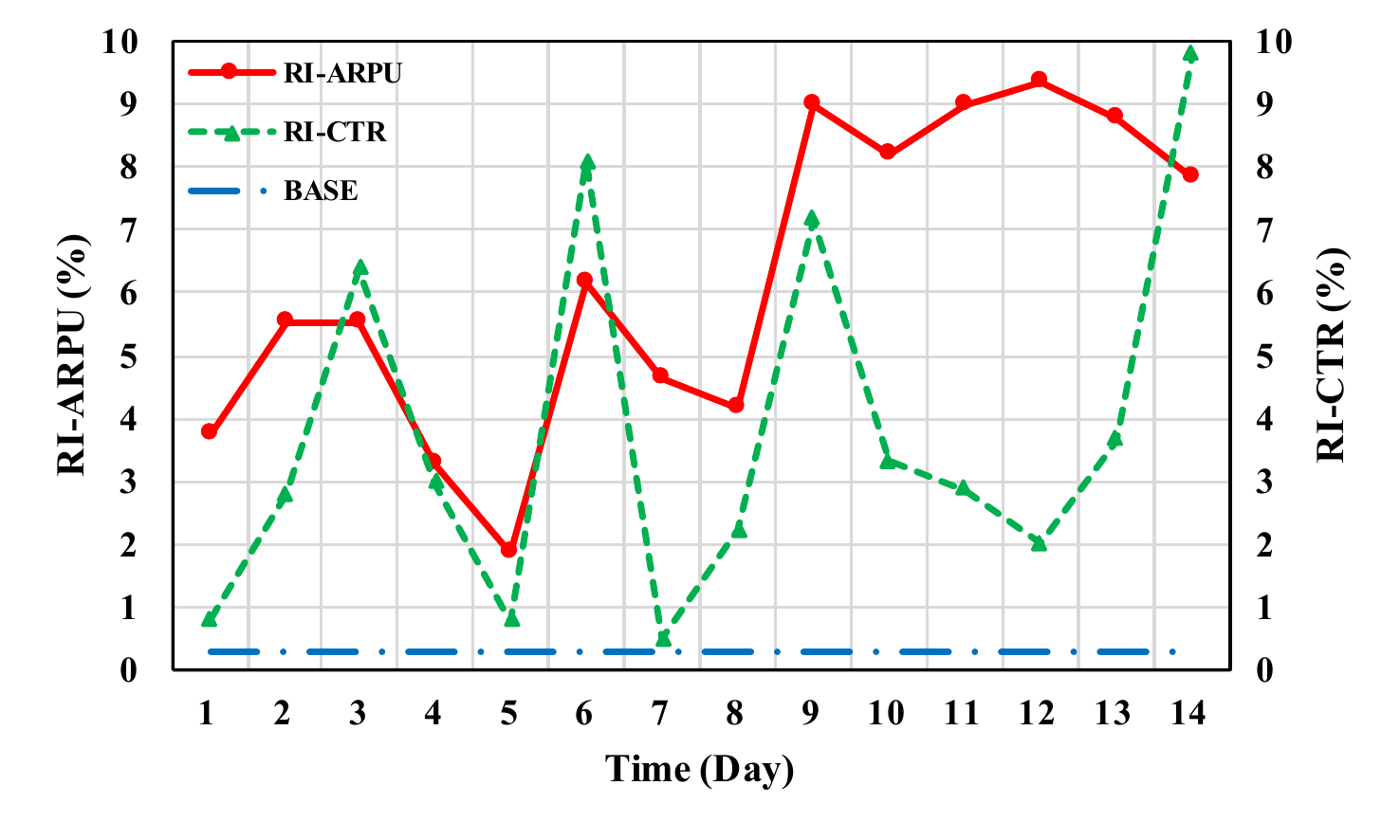}
\caption{Online evaluation of the proposed TFNet model against the base model (Wide\&Deep). }
\label{online}
\end{figure}

For the online advertising system, there are usually around $10\%$  new records of ads in the database each day. Therefore, it is necessary to consider new-comers of ads and users each day, which is the main difference of online and offline evaluations \cite{chan2018convolutional}. Moreover, the unstable online data distribution will lead to fluctuation of prediction performance, which can be illustrated in Figure \ref{online}. During the period of A/B tests, we calculate RI-ARPU and RI-CTR of the TFNet model each day. Great improvements of the TFNet model are observed: an average of 6.22\% relative ARPU improvement (max 9.34\%, min 1.88\%) and an average of 3.81\% relative CTR improvement (max 9.77\%, min 0.50\%). As is illustrated by the Tencent MyAPP, the budget of the latency in online system is $50ms$ at most, thus our proposed TFNet model's $8$$\sim$$17ms$ quite satisfies the common budget.

\section{Conclusion and Future Work}
In this work, we propose a tensor-based feature interaction model TFNet, which can learn the feature interactions in different semantic spaces. Extensive offline experiments demonstrate that the TFNet model significantly outperforms existing models and achieves the state-of-the-art results. Moreover, online A/B tests show great revenue and CTR improvements of the TFNet model in the largest Chinese App recommender system.

In addition, we are trying to utilize this model for ranking on the Chinese mainstream short video platform WeSee\footnote{https://www.weishi.com}, where the scene is a sliding play style instead of a click one, making it more complicated and challenging for samples modeling. It is in the offline verification stage at present, and online evaluation will be accessed later.

\balance  


%

\bibliographystyle{ACM-Reference-Format}
\bibliography{sigir20}


\end{document}